\documentclass[
aps,%
12pt,%
final,%
notitlepage,%
oneside,%
onecolumn,%
nobibnotes,%
nofootinbib,%
superscriptaddress,%
showpacs,%
centertags]%
{revtex4}
\def\bb {\begin {eqnarray}}
\def\ee {\end {eqnarray}}

\usepackage{rotating}

\usepackage{epsfig}
\usepackage{bm}% bold math

\def \ms {{\overline{\mbox{MS}}}}

\begin{document}
%\selectlanguage{english}
\title
{%Small-x behavior 
$Q^2$-evolution of parton densities at small-$x$
%low Q2 
values}
\author{\firstname{A.Yu.}~\surname{Illarionov}}
\affiliation{Dipartimento di Fisica dell'Universit\'{a} di Trento,
via Sommarive 14, I--38050 Povo, Trento, Italy.\\ E-mail: 
illario@science.unitn.it}
\author{\firstname{A.V.}~\surname{Kotikov}}
\affiliation{Joint Institute for Nuclear Research, Dubna, Russia.\\
E-mail: kotikov@theor.jinr.ru}

\begin{abstract}
In the
leading twist approximation of the Wilson operator product expansion
with ``frozen'' and analytic 
strong coupling constants we show that
Bessel-inspired
behavior of 
the structure function $F_2$ at small $x$,
obtained for
a flat initial condition in the DGLAP evolution equations,
leads to 
good agreement with the 
deep inelastic 
scattering  experimental data from HERA.

\end{abstract}

\maketitle

\section{Introduction}

The experimental data from
HERA on the deep-inelastic scattering (DIS) structure function
(SF) $F_2$ \cite{H197,ZEUS01},
and its derivatives $\partial F_2/\partial \ln(Q^2)$ \cite{H197,Surrow} and 
$\partial \ln F_2/\partial \ln(1/x)$ \cite{H1slo,Surrow,DIS02} enter us into
a very interesting kinematical range for
testing the theoretical ideas on the behavior of quarks and gluons carrying
a very low fraction of momentum of the proton, the so-called small-$x$
region. In this limit one expects that 
the conventional treatment based on the
Dokshitzer--Gribov--Lipatov--Altarelli--Parisi (DGLAP) equations
does not account for contributions to the cross section which are
leading in $\alpha_s \ln(1/x)$ and, moreover, the parton densities (PD), in 
particular the gluons, are becoming large and need to develop a high density 
formulation of QCD.

 However, the
reasonable agreement between HERA data and the next-to-leading-order (NLO) approximation of
perturbative
QCD has been observed for $Q^2 \geq 2$ GeV$^2$ (see reviews in \cite{CoDeRo}
and references therein) and, thus,
perturbative QCD could describe the
evolution of $F_2$ and its derivatives
%structure functions
up to very low $Q^2$ values,
traditionally explained by soft processes.
%It is of fundamental importance to find out the kinematical region where
%the well-established perturbative QCD formalism
%can be safely applied at small $x$.

The standard program to study the $x$ behavior of
quarks and gluons
is carried out by comparison of data
with the numerical solution of the DGLAP
%Dokshitzer-Gribov-Lipatov-Altarelli-Parisi (DGLAP)
equations 
\cite{DGLAP}\footnote{ At small $x$ there is another approach
based on the Balitsky--Fadin--Kuraev--Lipatov (BFKL) equation 
\cite{BFKL}, whose
application is out of the scope of this work. } by
fitting the parameters of the
$x$-profile of partons at some initial $Q_0^2$ and
the QCD energy scale $\Lambda$ \cite{fits,GRV,Ourfits}.
However, for analyzing exclusively the
small-$x$ region, there is the alternative of doing a simpler analysis
by using some of the existing analytical solutions of DGLAP 
in the small-$x$
limit \cite{BF1}--\cite{HT}.
This was done so in \cite{BF1}
where it was pointed out that the HERA small-$x$ data can be
interpreted in 
terms of the so-called doubled asymptotic scaling (DAS) phenomenon
related to the asymptotic 
behavior of the DGLAP evolution 
discovered many years ago \cite{Rujula}.

The study of \cite{BF1} was extended in \cite{Munich,Q2evo,HT}
to include the finite parts of anomalous dimensions
of Wilson operators and Wilson coefficients\footnote{ 
In the standard DAS approximation \cite{Rujula} only the singular
parts of the anomalous dimensions were used.}.
This has led to predictions \cite{Q2evo,HT} of the small-$x$ asymptotic PD
form 
%of parton distributions (PD)
in the framework of the DGLAP dynamics
%equation 
starting at some $Q^2_0$ with
the flat function
 \begin{eqnarray}
f_a (Q^2_0) ~=~
A_a ~~~~(\mbox{hereafter } a=q,g), \label{1}
 \end{eqnarray}
where $f_a$ are the parton distributions multiplied by $x$
and $A_a$ are unknown parameters to be determined from the data.

%From now on, we 
We refer to the approach of \cite{Munich,Q2evo,HT} as
{\it generalized} DAS approximation. In the approach
%generalized DAS 
the flat initial conditions in Eq. (\ref{1}) determine the
basic role of the singular parts of anomalous dimensions,
as in the standard DAS case, while
the contribution from finite parts 
of anomalous dimensions and from Wilson coefficients can be
considered as corrections which are, however, important for better 
agreement with experimental data.
In the present paper, similary to
\cite{BF1}--\cite{HT}, we neglect
the contribution from the non-singlet quark component.

The use of the flat initial condition given in Eq. (\ref{1}) is
supported by the actual experimental situation: low-$Q^2$ data
\cite{NMC,H197,lowQ2,Surrow} are well described for $Q^2 \leq 0.4$ GeV$^2$
by Regge theory with Pomeron intercept
$\alpha_P(0) \equiv \lambda_P +1 =1.08$,
closed to the standard ($\alpha_P(0) =1$) one. 
The small rise of HERA data \cite{H197,Surrow,lowQ2,lowQ2N}
at low $Q^2$ can be 
%naturally 
explained, for example, by contributions of
%including
higher twist operators
%terms 
(see \cite{HT}).
%Moreover, HERA data \cite{H197,H101,lowQ2,lowQ2N}
%with $Q^2 > 1$ GeV$^2$ are in good agreement with predictions from
%the GRV parton densities \cite{GRV}, 
%which 
%are very closed, at least conceptually, to the generalized DAS approach.\\

The purpose of this paper is to demostrate a good agreement between 
the predictions from the generalized DAS approach and the HERA experimental data
\cite{H197,ZEUS01} (see Figs. 1--3) for SF $F_2$  and also
to compare the predictions 
%from the generalized DAS approach 
for
%the small $Q^2$ behavior of 
the slope $\partial  \ln F_2/\partial \ln (1/x)$ 
%of the structure function $F_2$
with the 
%modern 
H1 and ZEUS 
%experimental 
data \cite{H1slo,Surrow,DIS02}.
%[H1slope,ZEUSslope].
%Very recently new precise experimental data on $\lambda (Q^2)$ 
%has become available \cite{H1slo}.
Looking at the H1 data \cite{H197} points shown in Figs. 4 and 5 one can conclude 
that $\lambda (Q^2)$
is independent on $x$ within the experimental uncertainties
for fixed $Q^2$ in the range $x <0.01$. Indeed, the data are well
described by the power behavior
\begin{eqnarray}
F_2 (x,Q^2) ~=~ C x^{-\lambda (Q^2)},
\label{1dd}
\end{eqnarray}
where $\lambda (Q^2) = \hat a \ln(Q^2/\Lambda^2)$ with $C \approx 0.18,
\hat a \approx 0.048$ and $ \Lambda =292$ MeV \cite{H1slo}.
The linear rise of the exponent $\lambda (Q^2)$ with $\ln Q^2$ is also 
explicitly shown in Figs. 4 and 5 by the dashed line.
%curve.

The
rise of $\lambda (Q^2)$ linearly with $\ln Q^2$ can be tracted in strong 
nonperturbative way (see \cite{Schrempp} and references therein), i.e.,
$\lambda (Q^2) \sim 1/\alpha_s(Q^2)$. The previous analysis \cite{KoPa02}, 
however, demonstrated
that the rise can be explained naturally in the framework of 
perturbative QCD.

%A similar analysis extracting $\lambda (Q^2)$ as a function of $x$
%has been carried out by the ZEUS Collaboration. 
%This behavior can also be inferred
%\footnote{ 
%These points lie slightly below the corresponding ZEUS data but all the
%results are in agreement within modern experimental errors.}
%from Fig. 2.\\

The ZEUS and H1 Collaborations have also presented \cite{Surrow,DIS02} 
%several
new preliminary data
%points 
for $\lambda (Q^2)$ at quite low values of $Q^2$.
As it
is possible to see in Fig. 8 of  \cite{Surrow}, the ZEUS value for 
$\lambda (Q^2)$ is consistent with a constant $\sim 0.1$ at $Q^2 <
0.6$ GeV$^2$, as it is expected under the assumption of single soft Pomeron
exchange within the framework of Regge phenomenology. 
These points lie slightly below the corresponding ZEUS data but all the
results are in agreement within modern experimental errors.

It is important to extend the analysis of \cite{KoPa02} to low $Q^2$ range
with a help of well-known infrared modifications of the strong coupling 
constant. We will use the ``frozen'' and analytic versions (see,  
\cite{Greco} and \cite{ShiSo}, respectively).

The paper is organized as follows.  Section 2 contains 
basic formulae,
%for the slope $d \ln F_2/d\ln (1/x)$ in generalized DAS
%%double-logarithmic approximation, 
which are
needed for the present study and were
%that were
previously obtained in \cite{Q2evo,HT,KoPa02}.
In Section 3 we compare our calculations with H1 and ZEUS
%$d\ln F_2/d\ln(1/x)$
experimental data and present
%discuss 
the obtained results.
Some discussions can be found in Conclusions.

\section{
generalized DAS
approach} \indent

The flat initial condition (\ref{1}) corresponds to the case when parton distributions
tend  to some constant value at $x \to 0$ and at some initial value $Q^2_0$.
%(\ref{1}).
The main ingredients of the results \cite{Q2evo,HT}, are:
\begin{itemize}
\item
Both, the gluon and quark singlet densities are presented in terms of two
components ($"+"$ and $"-"$) which are obtained from the analytic 
$Q^2$-dependent expressions of the corresponding ($"+"$ and $"-"$) PD
%parton distributions 
moments.
\item
The twist-two part of the $"-"$ component is constant at small $x$ at any 
values of $Q^2$,
whereas the one of the $"+"$ component grows at $Q^2 \geq Q^2_0$ as
\begin{equation}
\sim e^{\sigma},~~~
%\exp{\sigma},~~~
\sigma = 2\sqrt{\left[ \hat{d}_+ s
%\ln \left( \frac{a_s(Q^2_0)}{a_s(Q^2)} \right) 
- \left( \hat{D}_+ +  \hat{d}_+ \frac{\beta_1}{\beta_0} \right) p
% \Bigl( a_s(Q^2_0) - a_s(Q^2) \Bigr)
\right] \ln \left( \frac{1}{x} \right)}  \ ,~~~ \rho=\frac{\sigma}{2\ln(1/x)} \ ,
\label{intro:1}
\end{equation}
where $\sigma$ and $\rho$
%$=\sigma/(2\ln(1/x))$ 
are the generalized Ball--Forte
variables,
\begin{equation}
s=\ln \left( \frac{a_s(Q^2_0)}{a_s(Q^2)} \right),~~
p= a_s(Q^2_0) - a_s(Q^2),~~~
\hat{d}_+ = \frac{12}{\beta_0},~~~
\hat{D}_+ =  \frac{412}{27\beta_0}.
\label{intro:1a}
\end{equation}
%$8[23 C_A - 26 C_F]T_Rf/(9\beta_0)$.
%%\item
%%The recently observed difference between small $x$ behavior of sea
%%quark and gluon densities at $Q^2=Q^2_0$ are incorporated (in [IKP])
%%by high-twist corrections to the above twist-two approximation.
\end{itemize}
Hereafter we use the notation
$a_s=\alpha_s/(4\pi)$.
The first two coefficients of the QCD $\beta$-function in the $\ms$-scheme
are $\beta_0 = 11 -(2/3) f$
%$(11/3) C_A - (4/3) T_R f$ 
and $\beta_1 =  102 -(114/9) f$
%$(2/3)[17 C_A^2 - 10 C_A T_R f - 6 C_F T_R f]$ 
with $f$ is being the number of active quark flavors.
%This new presentation as a function of the
%$SU(N)$ group casimirs, with $f$ active flavors, $C_A = N$, $T_R = 1/2$,
%$T_F = T_R f$ and  $C_F = (N^2 - 1)/(2N)$ permits to apply our results
%to,  for example, the popular $N=1$ supersymmetric model.
%Of course, for $N=3$ one has the QCD result.
% \cite{Kotikov:1999}.

Note here that the perturbative coupling constant $a_s(Q^2)$ is different at
the leading-order (LO) and NLO approximations. Indeed, from the renormalization group equation
we can obtain the following equations for the coupling constant
%\begin{subequations}
%\label{as:LO&NLO}
\begin{eqnarray}
 \frac{1}{a_s(Q^2)} \, = \, \beta_0 
 \ln{\left(\frac{Q^2}{\Lambda^2_{\rm LO}}\right)}
\label{as:LO} 
\end{eqnarray}
at the LO approximation and
\begin{eqnarray}
 \frac{1}{a_s(Q^2)} \, + \,
 \frac{\beta_1}{\beta_0} \ln{\left[
 \frac{\beta_0^2 a_s(Q^2)}{\beta_0+ \beta_1 a_s(Q^2)}\right]} \, = \, 
 \beta_0 \ln{\left(\frac{Q^2}{\Lambda^2}\right)}
\label{as:NLO}
\end{eqnarray}
at the NLO approximation.
%\end{subequations}
Usually at the NLO level ${\rm \overline{MS}}$-scheme is used, so we apply
$\Lambda = \Lambda_{\rm \overline{MS}}$ below.
%in the Eqs.~(\ref{an:NLO}) and (\ref{as:NLO}).

\subsection{Parton distributions and the structure function $F_2$
%$Q^2$ dependence of the slope $d \ln F_2/d\ln (1/x)$ in g
%Generalized DAS approach
%double-logarithmic approximation
} 

%In this subsection
Here, for simplicity we consider only  the LO
%leading order (LO) 
approximation\footnote{
The NLO results may be found in  \cite{Q2evo,HT}.}.
The results for parton densities and $F_2$
%of Refs. \cite{Q2evo,HT} 
are following:
\begin{itemize}
\item
The structure function $F_2$ has the form
%Both, the gluon and quark singlet densities are presented in terms of two
%components ($'+'$ and $'-'$) 
\begin{eqnarray}
	F_2(x,Q^2) &=& e \, f_q(x,Q^2),
%\label{r10} 
\nonumber \\
	f_a(x,Q^2) &=& f_a^{+}(x,Q^2) + f_a^{-}(x,Q^2),
\label{8a}
\end{eqnarray}
where
%which are obtained from the analytic 
%$Q^2$-dependent expressions of the corresponding ($'+'$ and $'-'$) PDF
%%parton distributions 
%moments. Here, 
$e=(\sum_1^f e_i^2)/f$ is the average charge square.
%and $f$ is the number of active quark flavors.
\item
The small-$x$ asymptotic results for PD $f^{\pm}_a$ are
\begin{eqnarray}
	f^{+}_g(x,Q^2) &=& \biggl(A_g + \frac{4}{9} A_q \biggl)
		\tilde I_0(\sigma) \; e^{-\overline d_{+}(1) s} ~+~ O(\rho),
	\label{8.0} \\
	f^{+}_q(x,Q^2) &=& \frac{f}{9}\biggl(A_g + \frac{4}{9} A_q \biggl)
		\rho \tilde I_1(\sigma) \; e^{-\overline d_{+}(1) s} ~+~ O(\rho),
	\label{8.01} \\
	f^{-}_g(x,Q^2) &=& -\frac{4}{9} A_q e^{- d_{-}(1) s} ~+~ O(x),
	\label{8.00} \\
	f^{-}_q(x,Q^2) &=& A_q e^{-d_{-}(1) s} ~+~ O(x),
	\label{8.02}
\end{eqnarray}
where
$\overline d_{+}(1) = 1 + 20f/(27\beta_0)$ and
$          d_{-}(1) = 16f/(27\beta_0)$
are the regular parts of the anomalous dimensions $d_{+}(n)$ and $d_{-}(n)$, 
respectively, in the limit $n\to1$\footnote{
We denote the singular and regular parts of a given quantity $k(n)$ in the
limit $n\to1$ by $\hat k(n)$ and $\overline k(n)$, respectively.}.
%Here $n$ is the variable of the PD Mellin transform.
Here $n$ is
the variable in Mellin space.
%
%We define the variable
% \begin{eqnarray}
%s=ln\left(\frac{a_s(Q^2_0)}{a_s(Q^2)}\right)
%\label{2.4}
% \end{eqnarray}
%
The functions $\tilde I_{\nu}$ ($\nu=0,1$) 
%in Eqs. (\ref{8.0,8.01}) 
are related to the modified Bessel
function $I_{\nu}$
and to the Bessel function $J_{\nu}$ by:
\begin{equation}
\tilde I_{\nu}(\sigma) =
\left\{
\begin{array}{ll}
I_{\nu}(\sigma), & \mbox{ if } s \geq 0 \\
i^{-\nu} J_{\nu}(i\sigma), \ i^2=-1, \ & \mbox{ if } s \leq 0 
\end{array}
\right. .
\label{4}
\end{equation}
At the LO, 
the variables 
%$s$, 
$\sigma$ and $\rho$ are
%argument $\sigma$ is 
given by Eq. (\ref{intro:1}) when $p=0$.
\end{itemize}

\subsection{Effective slopes}

Contrary to the approach in  \cite{BF1,Munich,Q2evo,HT}
%\cite{BF1}-\cite{HT},
%On the other hand, 
various groups have been able to fit
the available data 
%({\it separately at low and high $Q^2$ values})
using a hard input at small $x$: 
$x^{-\lambda},~\lambda >0$ with different $\lambda$ values at low and high 
$Q^2$ (see \cite{LoYn,DoLa,Abramo,YF93,FKR,CaKaMeTTV}).
Such results are well-known at low $Q^2$ values \cite{DoLa}. 
At large $Q^2$ values, for 
%In some sense, it is not very surprising, because 
the modern HERA data 
it is also not very surprising, because they
cannot distinguish between the behavior
based on a steep input parton parameterization,
at quite large $Q^2$, and the
steep form acquired after the dynamical evolution from a flat initial
condition at quite low $Q^2$ values.

As it has been mentioned above and shown in \cite{Q2evo,HT},
the behavior of parton densities and $F_2$ given in the Bessel-like form 
by generalized DAS approach
%Eqs. (\ref{9.10})-(\ref{9}) 
can mimic a power law shape
over a limited region of $x$ and $Q^2$
 \begin{eqnarray}
f_a(x,Q^2) \sim x^{-\lambda^{\rm eff}_a(x,Q^2)}
 ~\mbox{ and }~
F_2(x,Q^2) \sim x^{-\lambda^{\rm eff}_{\rm F_2}(x,Q^2)}.
\nonumber    \end{eqnarray}

The effective slopes $\lambda^{\rm eff}_a(x,Q^2)$ and $\lambda^{\rm eff}_{\rm F_2}(x,Q^2)$
have the form:
 \begin{eqnarray}
\lambda^{\rm eff}_g(x,Q^2) &=& \frac{f^+_g(x,Q^2)}{f_g(x,Q^2)} \,
\rho \, \frac{\tilde I_1(\sigma)}{\tilde I_0(\sigma)},
\nonumber
\\
\lambda^{\rm eff}_q(x,Q^2) &=& \frac{f^+_q(x,Q^2)}{f_q(x,Q^2)} \,
\rho \, \frac{\tilde  I_2(\sigma) (1- 20 a_s(Q^2))
 + 20 a_s(Q^2) \tilde I_1(\sigma)/\rho}{\tilde  I_1(\sigma) 
(1- 20 a_s(Q^2))
 + 20 a_s(Q^2) \tilde I_0(\sigma)/\rho},
\nonumber
%\label{10.1}
%
\\
\lambda^{\rm eff}_{\rm F_2}(x,Q^2) &=& \frac{\lambda^{eff}_q(x,Q^2) \,
f^+_q(x,Q^2) + (2f)/3a_s(Q^2)\, \lambda^{eff}_g(x,Q^2) \,
f^+_g(x,Q^2)}{f_q(x,Q^2) + (2f)/3 a_s(Q^2)\, f_g(x,Q^2)},
%\nonumber
\label{10.1}
\end{eqnarray}
where 
the exact form of parton densities can be found in \cite{Q2evo,HT}.

The results (\ref{10.1}) (and also (\ref{11.1a})--(\ref{11.1}) below) are 
given at the
NLO approximation. To obtain the LO
one, it is necessary to cancel the term $\sim a_s(Q^2)$ and to use Eqs. 
(\ref{8.0})--(\ref{8.02}) for parton densities $f_a(x,Q^2)$.

The effective slopes $\lambda^{\rm eff}_a $ and 
$\lambda^{\rm eff}_{\rm F_2}$ depend on the magnitudes $A_a$ of the initial PD
and also on the chosen input values of $Q^2_0$ and $\Lambda $.
To compare with the experimental data it is necessary the exact expressions
(\ref{10.1}), but for qualitative analysis it is better to use an 
approximation.

\subsection{Asymptotic form of the effective slopes}

At quite 
large values of $Q^2$, 
where the ``$-$'' component is negligible,
the dependence on the initial PD disappears, having
in this case for the asymptotic behavior the following 
expressions\footnote{The asymptotic formulae given in 
Eqs. (\ref{11.1a})--(\ref{11.1})
work quite well at any $Q^2 \geq Q^2_0$ values,
because at $Q^2=Q^2_0$ the values of
$\lambda^{\rm eff}_a $ and $\lambda^{\rm eff}_{\rm F_2}$ are equal zero. 
The use of approximations in Eqs. (\ref{11.1a})--(\ref{11.1}) instead of the exact results 
given in Eq. (\ref{10.1}) underestimates 
(overestimates) only slightly the gluon (quark) slope
at $Q^2 \geq Q^2_0$.
%For the $F_2$ case, the similarity of $\lambda^{eff}_{F2} $ and 
%$\lambda^{eff,as}_{F2}$ values is shown in Fig 1.
}:
 \begin{eqnarray}
\lambda^{\rm eff,as}_g(x,Q^2) &=& 
\rho\, \frac{\tilde I_1(\sigma)}{\tilde I_0(\sigma)} \approx \rho - 
\frac{1}{4\ln{(1/x)}}, 
%\nonumber \\
\label{11.1a} \\
\lambda^{\rm eff,as}_q(x,Q^2) &=& 
\rho \frac{\tilde  I_2(\sigma) (1- 20 a_s(Q^2))
 + 20 a_s(Q^2)\tilde  I_1(\sigma)/\rho}{\tilde  I_1(\sigma) 
(1- 20 a_s(Q^2))
 + 20 a_s(Q^2)\tilde  I_0(\sigma)/\rho}
 \nonumber \\
%&=& \rho\, \frac{\tilde I_2(\sigma)}{\tilde I_1(\sigma)} 
% + 20 a_s(Q^2)\left( 1- \frac{\tilde  I_0(\sigma) 
%\tilde  I_2(\sigma)}{\tilde  I_1^2(\sigma)} \right)
&\approx &
 \rho - \frac{3}{4\ln{(1/x)}} +  
\frac{10a_s(Q^2)}{ \rho \ln{(1/x)}}, 
%
% \nonumber \\
\label{11.1b}\\ 
\lambda^{\rm eff,as}_{\rm F_2}(x,Q^2) 
&=& \rho\, \frac{\tilde I_2(\sigma)}{\tilde I_1(\sigma)} 
 + 26 a_s(Q^2)\left( 1- \frac{\tilde  I_0(\sigma) 
\tilde  I_2(\sigma)}{\tilde  I_1^2(\sigma)} \right)
 \nonumber \\
%&=& \lambda^{\rm eff,as}_q(x,Q^2) + 6 a_s(Q^2)\left( 1- \frac{\tilde  I_0(\sigma) 
%\tilde  I_2(\sigma)}{\tilde  I_1^2(\sigma)} \right)
% \nonumber \\
&\approx &
 \rho - \frac{3}{4\ln{(1/x)}} +  
\frac{13a_s(Q^2)}{ \rho \ln{(1/x)}} 
= \lambda^{\rm eff,as}_q(x,Q^2)  +  
\frac{3a_s(Q^2)}{ \rho \ln{(1/x)}},
%\nonumber
\label{11.1} 
\end{eqnarray}
where the symbol $\approx $ marks the approximation obtained in the  expansion
of the usual and modified Bessel functions in (\ref{4}).
% $I_n(\sigma)$ $(n=0,1,2)$. 
These approximations are
accurate only at very large $\sigma $ values (i.e. at very large $Q^2$
and/or very small $x$).

As one can see from Eqs. (\ref{11.1a}) and (\ref{11.1b}),  
%As it has been already shown in Ref. \cite{Q2evo},
the gluon effective slope $\lambda^{\rm eff}_g$ 
is larger than the quark slope
$\lambda^{\rm eff}_q$, which is in excellent agreement with 
%a recent
MRS \cite{MRS} and GRV \cite{GRV} analyses (see also \cite{fits}).

We would like to note that at the NLO approximation the slope 
$\lambda^{\rm eff,as}_{\rm F_2}(x,Q^2)$ lies between quark and
gluon ones but closely to quark slope 
$\lambda^{\rm eff,as}_{q}(x,Q^2)$.
Indeed,
 \begin{eqnarray}
 \lambda^{\rm eff,as}_{g}(x, Q^2) \ - \
 \lambda^{\rm eff,as}_{\rm F_2}(x, Q^2) \ 
%&=& \
% \left(\rho \, \frac{\widetilde{I}_1(\sigma)}{\widetilde{I}_0(\sigma)}
% + 26 a_s(Q^2) \right)
%\left(1 -
%   \frac{\widetilde{I}_0(\sigma) \widetilde{I}_2(\sigma)}
%         {\widetilde{I}_1^2(\sigma)} \right)
%\nonumber \\
 \ &\approx& \ \left(\rho - \frac{1}{4\ln{(1/x)}}
 + 26 a_s(Q^2)\right) \frac{1}{2 \rho \ln{(1/x)}}  ,
\label{Slopes:NLO:G-F2} \\
 \lambda^{\rm eff,as}_{\rm F_2}(x, Q^2) \ - \
 \lambda^{\rm eff,as}_{q}(x, Q^2) \ 
%&=& \
% 6 \, a_s(Q^2) \left(1 -
%   \frac{\widetilde{I}_0(\sigma) \widetilde{I}_2(\sigma)}
 %        {\widetilde{I}_1^2(\sigma)} \right) \ \
&\approx &\
 \frac{3 a_s(Q^2)}{\rho \ln(1/x)}  .
\label{Slopes:NLO:F2-q}
 \end{eqnarray}

Both slopes $\lambda^{\rm eff}_a(x, Q^2)$ decrease with increasing $x$ (see Fig. 5). \
A $x$-dependence of the slope should not appear for a PD within a
Regge type asymptotic ($x^{-\lambda}$) and precise measurement of the slope 
$\lambda^{\rm eff}_a(x, Q^2)$ may lead to the possibility to verify the
type of small-$x$ asymptotics of parton distributions.

\section{Comparison with experimental data
%Results of the fits
} \indent

Using the results of previous section we have
analyzed  HERA data for $F_2$ and the slope $\partial \ln F_2/\partial \ln (1/x)$
at small $x$ from the H1 and ZEUS Collaborations \cite{H197,ZEUS01,Surrow,H1slo,DIS02}.

In order to keep the analysis as simple as possible,
we fix $f=4$ and $\alpha_s(M^2_Z)=0.1166 $ (i.e., $\Lambda^{(4)} = 284$ MeV) in agreement
with the more recent ZEUS results \cite{ZEUS01}.

As it is possible to see in Figs. 1--3 (see also \cite{Q2evo,HT}), the twist-two
approximation is reasonable at $Q^2 \geq 2$ GeV$^2$. At smaller $Q^2$, some
modification of the approximation should be considered. In the recent article
\cite{HT} we have added the higher twist corrections.
For renormalon model of higher twists, we
have found a good
agreement with experimental data at essentially lower $Q^2$ values:
$Q^2 \geq 0.5$ GeV$^2$ (see Figs. 2 and 3).

Moreover, 
 the results of fits in \cite{HT} have an important property: they are
very similar in LO and NLO approximations of perturbation theory.
The similarity is related to the fact that the small-$x$ asymptotics of 
the NLO corrections
are usually large and negative (see, for example, $\alpha_s$-corrections \cite{FaLi} to
BFKL kernel \cite{BFKL}\footnote{It seems that it is a property of 
any processes in which gluons,
but not quarks play a basic role.}).
% and 
Then, the LO form $\sim \alpha_s(Q^2)$ for
some observable and the NLO one 
$\sim \alpha_s(Q^2) (1-K\alpha_s(Q^2)) $
with a large value of $K$ are similar, because 
%usually 
$\Lambda \gg
\Lambda_{\rm LO}$\footnote{The equality of
%similarity between 
$\alpha_s(M_Z^2)$ at LO and NLO approximations,
%and $\alpha^{\rm LO}_s(M_Z^2)$,
where $M_Z$ is the $Z$-boson mass, relates $\Lambda$ and $\Lambda_{\rm LO}$:
$\Lambda^{(4)} = 284$ MeV (as in \cite{ZEUS01}) corresponds to 
$\Lambda_{\rm LO} = 112$ MeV (see \cite{HT}).}
and, thus, $\alpha_s(Q^2)$ at LO is considerably smaller  then 
$\alpha_s(Q^2)$ at NLO  for HERA $Q^2$ values.

In other words, performing some resummation procedure (such as Grunberg's 
effective-charge method \cite{Grunberg}), one can see that the 
NLO form may
%can 
be represented as $\sim \alpha_s(Q^2_{\rm eff})$,
where $Q^2_{\rm eff} \gg Q^2$. 
Indeed, from 
different studies
\cite{DoShi,bfklp,Andersson},
it is well known that at small-$x$ values the effective
argument of the coupling constant is higher then $Q^2$.

Here, to improve the agreement at small $Q^2$ values,
we modify the QCD coupling constant.
We consider two modifications, 
which effectively increase the argument of the coupling constant 
at small $Q^2$ values (in agreement with \cite{DoShi,bfklp,Andersson}).

In one case, which is more phenomenological, we introduce freezing
of the coupling constant by changing its argument $Q^2 \to Q^2 + M^2_{\rho}$,
where $M_{\rho}$ is the $\rho $-meson mass (see \cite{Greco}). Thus, in the 
formulae of the
Section 2 we should do the following replacement:
\begin{equation}
 a_s(Q^2) \to a_{\rm fr}(Q^2) \equiv a_s(Q^2 + M^2_{\rho})
\label{Intro:2}
\end{equation}

The second possibility incorporates the Shirkov--Solovtsov idea 
\cite{ShiSo,Nesterenko,Cvetic}
about analyticity of the coupling constant that leads to the additional its
power dependence. Then, in the formulae of the previous section
%and \ref{Sec:3}
the coupling constant $a_s(Q^2)$ should be replaced as follows:
\begin{eqnarray}
 a^{\rm LO}_{\rm an}(Q^2) \, = \, a_s(Q^2) - \frac{1}{\beta_0}
 \frac{\Lambda^2_{\rm LO}}{Q^2 - \Lambda^2_{\rm LO}}
\label{an:LO} 
\end{eqnarray}
at the LO
%leading (LO) 
approximation and
\begin{eqnarray}
 a_{\rm an}(Q^2) \, = \, a_s(Q^2) - \frac{1}{2\beta_0}
 \frac{\Lambda^2}{Q^2 - \Lambda^2} 
+ \ldots \, ,
%- \frac{1}{\beta_0}
% \sum_{k=1}^\infty \left(\frac{\Lambda^2}{Q^2}\right)^k \, C_k[f]
\label{an:NLO}
\end{eqnarray}
at the NLO approximation,
where the symbol $\ldots$ stands for terms which have negligible
contributions
at $Q \geq 1$ GeV \cite{ShiSo}\footnote{Note that in \cite{Nesterenko,Cvetic} 
more accurate, but essentially more
cumbersome approximations of $a_{an}(Q^2)$ have been proposed.
We limit ourselves by above simple form (\ref{an:LO}), (\ref{an:NLO})
and plan to add the other modifications in our future investigations.}.

Figure~4 shows the experimental data for $\lambda_{F_2}^{\rm eff}(x,Q^2)$
\footnote{Using the ``frozen'' and analytic coupling constants,
the experimental data for $F_2(x,Q^2)$ have been analysed recenly
in \cite{Cvetic1}.}
at $x\sim 10^{-3}$, which represents an average of the $x$-values of HERA experimental 
data. The top dashed line represents the aforementioned linear rise of
$\lambda(Q^2)$ with $\ln(Q^2)$.

%It 
Figure~4 demonstrates 
that the theoretical description of the small-$Q^2$ ZEUS
data for $\lambda^{\rm eff}_{F_2}(x,Q^2)$ by NLO QCD is significantly
improved by implementing the ``frozen'' and analytic coupling constants
$\alpha_{\rm fr}(Q^2)$ and $\alpha_{\rm an}(Q^2)$, 
respectively,
which in turn lead to
very close results (see also \cite{KoLiZo}).

Indeed, the fits for $F_2(x,Q^2)$ in \cite{HT}
yielded
$Q^2_0 \approx 0.5$--$0.8$~GeV$^2$.
So, initially we had $\lambda^{\rm eff}_{F_2}(x,Q^2_0)=0$,
as suggested by Eq.~(\ref{1}). The replacements of Eqs.~(\ref{Intro:2}), (\ref{an:LO}) 
and (\ref{an:NLO}) modify the
value of $\lambda^{\rm eff}_{F_2}(x,Q^2_0)$. 
%So, for 
For the  
``frozen'' and analytic coupling constants 
$\alpha_{\rm fr}(Q^2)$ and $\alpha_{\rm an}(Q^2)$,
%coupling constant $a_{fr}(Q^2)$ 
the value of
$\lambda^{\rm eff}_{F_2}(x,Q^2_0)$ is nonzero 
%now 
and the slopes are
%is 
quite close to the experimental data at $Q^2 \approx 0.5$~GeV$^2$.
Nevertheless, for $Q^2 \leq 0.5$~GeV$^2$, there is still some disagreement with
the data, which needs additional investigation.

Figure~5 shows the $x$-dependence of the slope 
$\lambda^{\rm eff}_{F_2}(x,Q^2)$.
One observes good agreement between the experimental data and the generalized
DAS approach for a broad range of small-$x$ values.
%One can see an 
The absence of a variation with $x$ of 
$\lambda^{\rm eff}_{F_2}(x,Q^2)$ at small $Q^2$ values is related to the small
values of the variable $\rho$ there.

At large $Q^2$ values, the $x$-dependence of $\lambda^{\rm eff}_{F_2}(x,Q^2)$ is
rather strong.
However, it is well known that
the boundaries and mean values of the experimental $x$ ranges 
\cite{H1slo} increase proportionally with $Q^2$, which is related
to the kinematical restrictions in the HERA experiments:
$x \sim 10^{-4} \times  Q^2$
(see \cite{H197,ZEUS01,KoPa02}
and, for example,
Fig. 1 of \cite{Surrow}).
We show only the case with the ``frozen'' coupling constant because at large $Q^2$ values
all results are very similar.

From Fig.~5, one can see that HERA experimental data are close to
$\lambda^{\rm eff}_{F_2}(x,Q^2)$ at $x \sim 10^{-4} \div 10^{-5}$
for $Q^2=4$~GeV$^2$ and at $x \sim 10^{-2}$ for $Q^2=100$~GeV$^2$. Indeed,
%Incorporation of the
the correlations between $x$ and $Q^2$  in the form 
$x_{\rm eff}= a \times 10^{-4} \times Q^2$ with $a=0.1$ and $1$
%and $10$, 
lead
%in our %the analysis 
to a modification of the $Q^2$ evolution which starts
to resemble $\ln Q^2$, rather than $\ln \ln Q^2$ as is standard 
\cite{KoPa02}.
%It is in agreement with H1 data \cite{H197} for 
%$a$ values between $0.1$ and $1$ and $Q^2 > 2$ GeV$^2$, which approximately
%corresponds to the middle points of the measured $x$ range.

\section{Conclusions} \indent

We shown
%studied 
the $Q^2$-dependence of the structure function $F_2$ and
the slope 
$\lambda^{\rm eff}_{F_2}=\partial \ln F_2/\partial \ln (1/x)$ at 
small-$x$ values in the 
framework of perturbative QCD. Our twist-two 
results are in very good agreement with 
%new 
precise HERA data at $Q^2 \geq 2$~GeV$^2$,
where perturbative theory can be applicable.
The application of the ``frozen'' and analytic coupling constants 
$\alpha_{\rm fr}(Q^2)$
and $\alpha_{\rm an}(Q^2)$ improves
%coupling constant $a_{fr}(Q^2)$
%(\ref{Intro:2}) 
%leads to good 
the agreement with the recent HERA data \cite{Surrow,H1slo,DIS02}
for the slope $\lambda^{\rm eff}_{F_2}(x,Q^2)$ for small $Q^2$ values,
$Q^2 \geq 0.5$~GeV$^2$.

As a next step of investigations, we plan to fit the HERA 
data \cite{H197,ZEUS01,Surrow,H1slo,DIS02}
of $F_2(x,Q^2)$  directly, using
% with 
the ``frozen'' and analytic 
coupling constants
%$\alpha_{fr}(Q^2)$ and $\alpha_{an}(Q^2)$
in both the LO and NLO approximations,
in order to improve the agreement with HERA data at small $Q^2$ values.
Several versions of the analytical coupling constant will be used.\\

%\indent
%\vspace{1cm} \hspace{1cm} {\Large {} {\bf Acknowledgments} %\vspace{0.5cm}
%}\\

\indent
This work was supported by 
RFBR grant 07-02-01046-a.
%Author 
A.V.K. thanks the Organizing Committee of 
Workshop on Physics of Fundamental Interactions
%, Institute of High Energy Physics, Protvino, Russia, 22--25 December 2008
%the conference of nuclear section of OFS RAS 
for invitation.

\newpage

\vspace{0.5cm}
%\vspace{2.5cm}

\hspace{1cm} {\Large{\bf Figure captions}}    \vspace{0.5cm}

{ \bf Fig. 1.} $F_2(x,Q^2)$ as a function of $x$ for different $Q^2$ bins. 
The experimental points are from H1 \cite{H197} (open points) and ZEUS 
\cite{ZEUS01} (solid points) at 
$Q^2 \geq 1.5$ GeV$^2$.
The solid curve represents the NLO fit. The dashed curve (hardly 
distinguishable 
from the solid one) represents the LO fit.

\vspace{0.5cm}

{ \bf Fig. 2.} $F_2(x,Q^2)$ as a function of $x$ for different $Q^2$ bins. 
The experimental points are same as on Fig. 1.
The solid curve represents the NLO fit. 
The dash-dotted curve represents the BFKL-motivated estimation for higher-twist
corrections to $F_2(x,Q^2)$ (see \cite{HT}).
The dashed curve is obtained from the fits at the NLO, when the renormalon contributions
of higher-twist terms have been incorporated.

\vspace{0.5cm}

{ \bf Fig. 3.} $F_2(x,Q^2)$ as a function of $x$ for different $Q^2$ bins. 
The experimental points are from H1 \cite{H197} 
  (open points) and ZEUS \cite{ZEUS01} (solid points)
at $Q^2 \geq 0.5$ GeV$^2$.
The solid curve represents the NLO fit. 
The dashed curve is from the fits at the NLO with the renormalon contributions
of higher-twist terms incorporated.
The dash-dotted curve (hardly distinguishable 
from the dashed one) represents the LO fit with the renormalon contributions
of higher-twist terms incorporated.

\vspace{0.5cm}

{ \bf Fig. 4.} The values of effective slope
$\lambda^{\rm eff}_{\rm F_2}$ 
as a function of $Q^2$ for $x=10^{-3}$.  
The experimental points are from H1 \cite{H1slo,DIS02} (open points) and ZEUS 
\cite{Surrow} (solid points). The solid curve represents the NLO fit.
The dash-dotted  and lower dashed curves represent the NLO fits with 
``frozen'' and analytic coupling constants, respectively. The top dashed line
 represents the fit from \cite{H1slo}.

\vspace{0.5cm}

{ \bf Figure 5.} The values of effective slope
$\lambda^{\rm eff}_{\rm F_2}$  
as a function of $Q^2$.
The experimental points are same as on Fig. 4.
The dashed line
 represents the fit from \cite{H1slo}.
The solid curves represent the NLO fits with 
``frozen'' coupling constant
at $x=10^{-2}$ and  $x=10^{-5}$.

\newpage

%

%  Fig. 1
\begin{figure}[t]
\begin{center}
  \setlength{\unitlength}{1.4mm}
\begin{picture}(125,125)   %figure:(120,77)
%    \coordsyst{120}{77}{25}{16}
 \put(5,5){
  \centering\epsfig{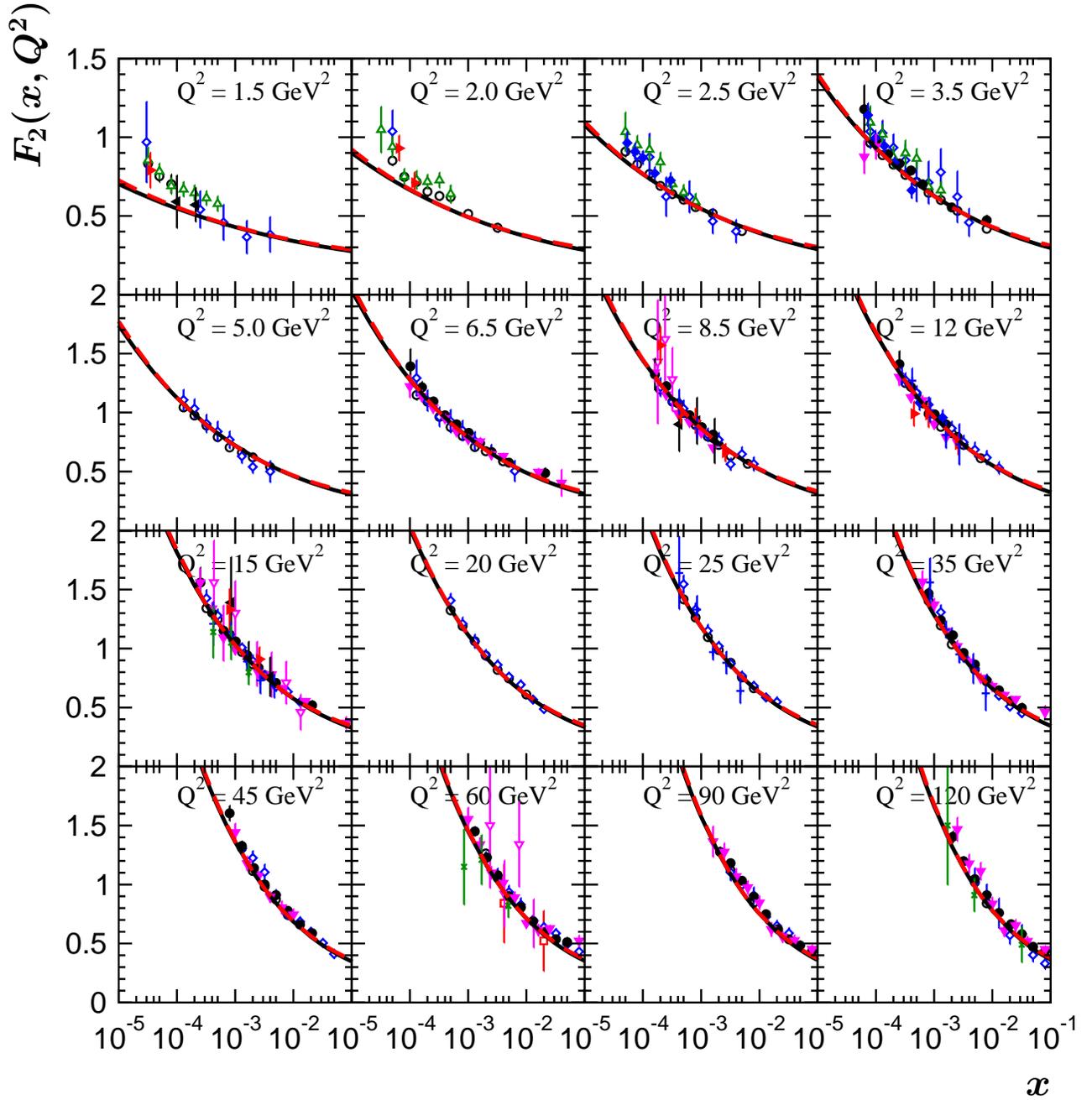}}
%  \centering\epsfig{file=F2-LOt2.eps, width=120\unitlength}}
%
 \put(115,0){\bfseries\Large
% \put(0,0){\bfseries\Large
 $\boldsymbol{x}$
%$x$
}
 \put(0,105){\bfseries\Large
% \put(0,135){\bfseries\Large
  \begin{sideways}
   $\boldsymbol{F_2(x, Q^2)}$
%$F_2(x, Q^2)$
  \end{sideways}}
\end{picture}
\end{center}
%\begin{figure}[htb]
%\begin{center}
%%\epsfig{figure= F2-LOt2.eps,width=17cm,height=13cm}
%\epsfig{figure= Fig1.eps,width=17cm,height=13cm}
%\end{center}
\vskip 1cm
\caption{$F_2(x,Q^2)$ as a function of $x$ for different $Q^2$ bins. 
The experimental points are from H1 \cite{H197} (open points) and ZEUS 
\cite{ZEUS01} (solid points) at 
$Q^2 \geq 1.5$ GeV$^2$.
The solid curve represents the NLO fit. The dashed curve (hardly 
distinguishable 
from the solid one) represents the LO fit.}
\label{fig1}
\end{figure}

%  Fig. 2
\begin{figure}[t]
\begin{center}
  \setlength{\unitlength}{1.4mm}
\begin{picture}(125,125)   %figure:(120,77)
%    \coordsyst{120}{77}{25}{16}
 \put(5,5){
%  \centering\epsfig{file=eps/LH1-LOt2.eps, width=120\unitlength}}
  \centering\epsfig{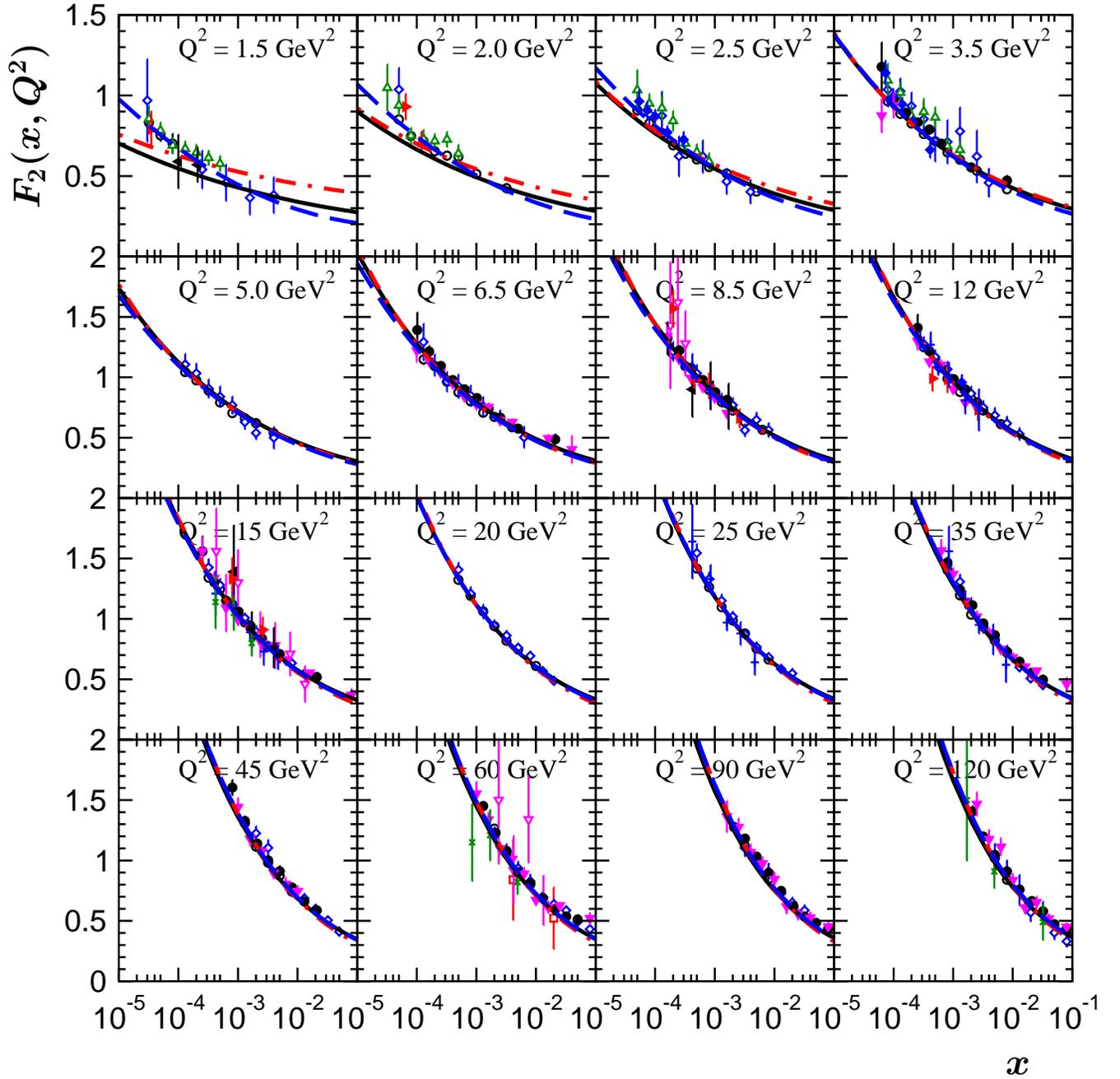}}
%  \centering\epsfig{file=LH1-LOt2.eps, width=120\unitlength}}
%
 \put(115,0){\bfseries\Large
  $\boldsymbol{x}$}
 \put(0,100){\bfseries\Large
  \begin{sideways}
%   $\boldsymbol{\lambda_{F_2}^{\rm eff}(x, Q^2)}$
$\boldsymbol{F_2(x, Q^2)}$
  \end{sideways}}
\end{picture}
\end{center}
%\begin{figure}[htb]
%\begin{center}
%%\epsfig{figure= F2-t2HT.eps,width=17cm,height=13cm}
%\epsfig{figure= Fig2.eps,width=17cm,height=13cm}
%\end{center}
\vskip 1cm
\caption{$F_2(x,Q^2)$ as a function of $x$ for different $Q^2$ bins. 
The experimental points are same as on Fig. 1.
The solid curve represents the NLO fit. 
The dash-dotted curve represents the BFKL-motivated estimation for higher-twist
corrections to $F_2(x,Q^2)$ (see \cite{HT}).
The dashed curve is obtained from the fits at the NLO, when the renormalon contributions
of higher-twist terms have been incorporated.}
\label{fig2}
\end{figure}

%  Fig. 3
\begin{figure}[t]
\begin{center}
  \setlength{\unitlength}{1.4mm}
\begin{picture}(125,125)   %figure:(120,77)
%    \coordsyst{120}{77}{25}{16}
 \put(5,5){
%  \centering\epsfig{file=eps/dF2_dQ2-m.eps, width=120\unitlength}}
  \centering\epsfig{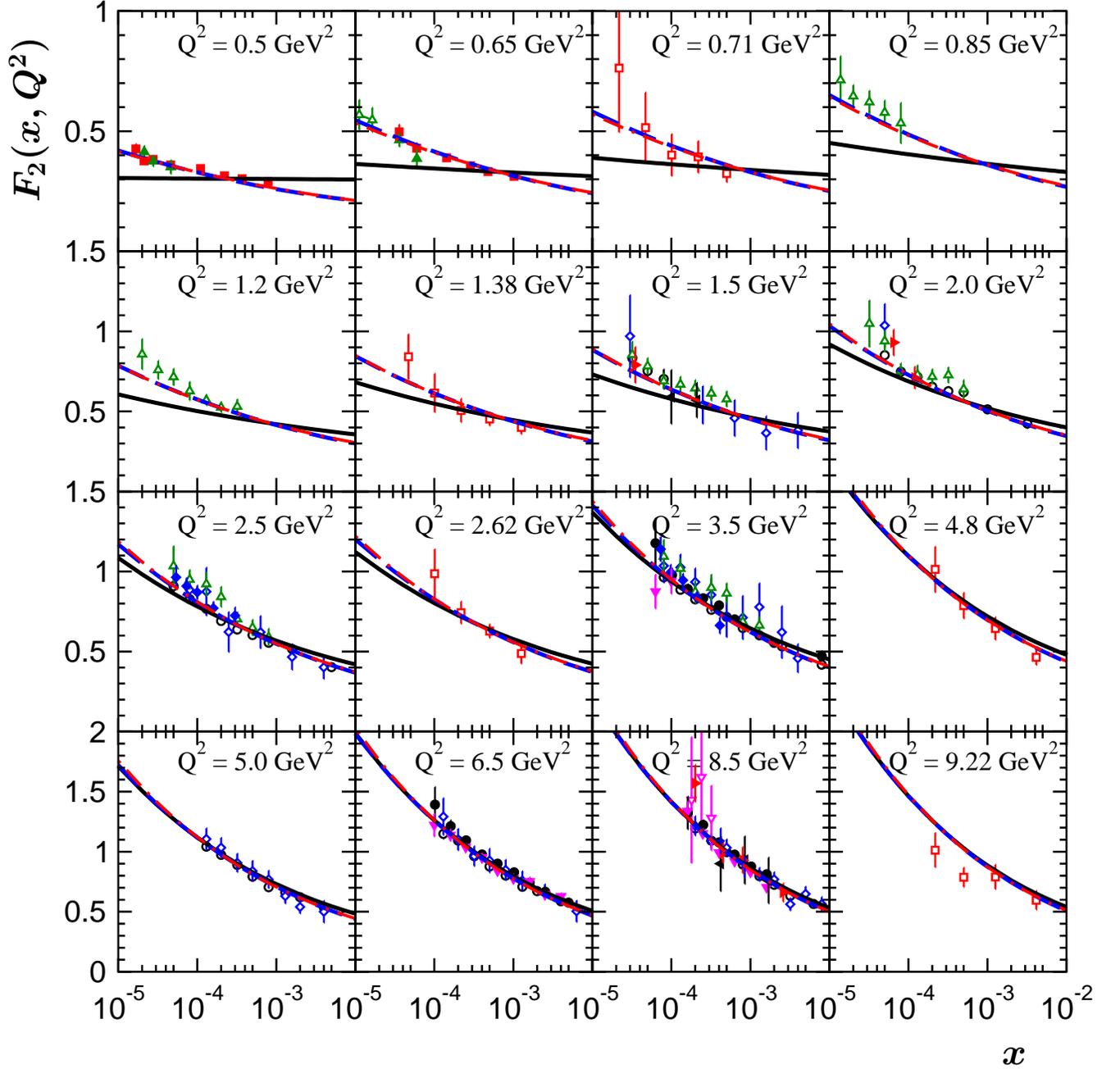}}
%  \centering\epsfig{file=dF2_dQ2-m.eps, width=120\unitlength}}
%
 \put(115,0){\bfseries\Large
  $\boldsymbol{x}$}
 \put(0,100){\bfseries\Large
  \begin{sideways}
$\boldsymbol{F_2(x, Q^2)}$
%   $\boldsymbol{\partial{F_2}/\partial\ln{Q^2}}$
  \end{sideways}}
\end{picture}
\end{center}
%\begin{figure}[htb]
%\begin{center}
%%\epsfig{figure= F2sm-t2HT.eps,width=17cm,height=13cm}
%\epsfig{figure= Fig3.eps,width=17cm,height=13cm}
%\end{center}
\vskip 1cm
\caption{$F_2(x,Q^2)$ as a function of $x$ for different $Q^2$ bins. 
The experimental points are from H1 \cite{H197} 
  (open points) and ZEUS \cite{ZEUS01} (solid points)
at $Q^2 \geq 0.5$ GeV$^2$.
The solid curve represents the NLO fit. 
The dashed curve is from the fits at the NLO with the renormalon contributions
of higher-twist terms incorporated.
The dash-dotted curve (hardly distinguishable 
from the dashed one) represents the LO fit with the renormalon contributions
of higher-twist terms incorporated.}
\label{fig3}
\end{figure}

\begin{figure}[htb]
\begin{center}
\includegraphics[height=4.9in,width=6.5in]{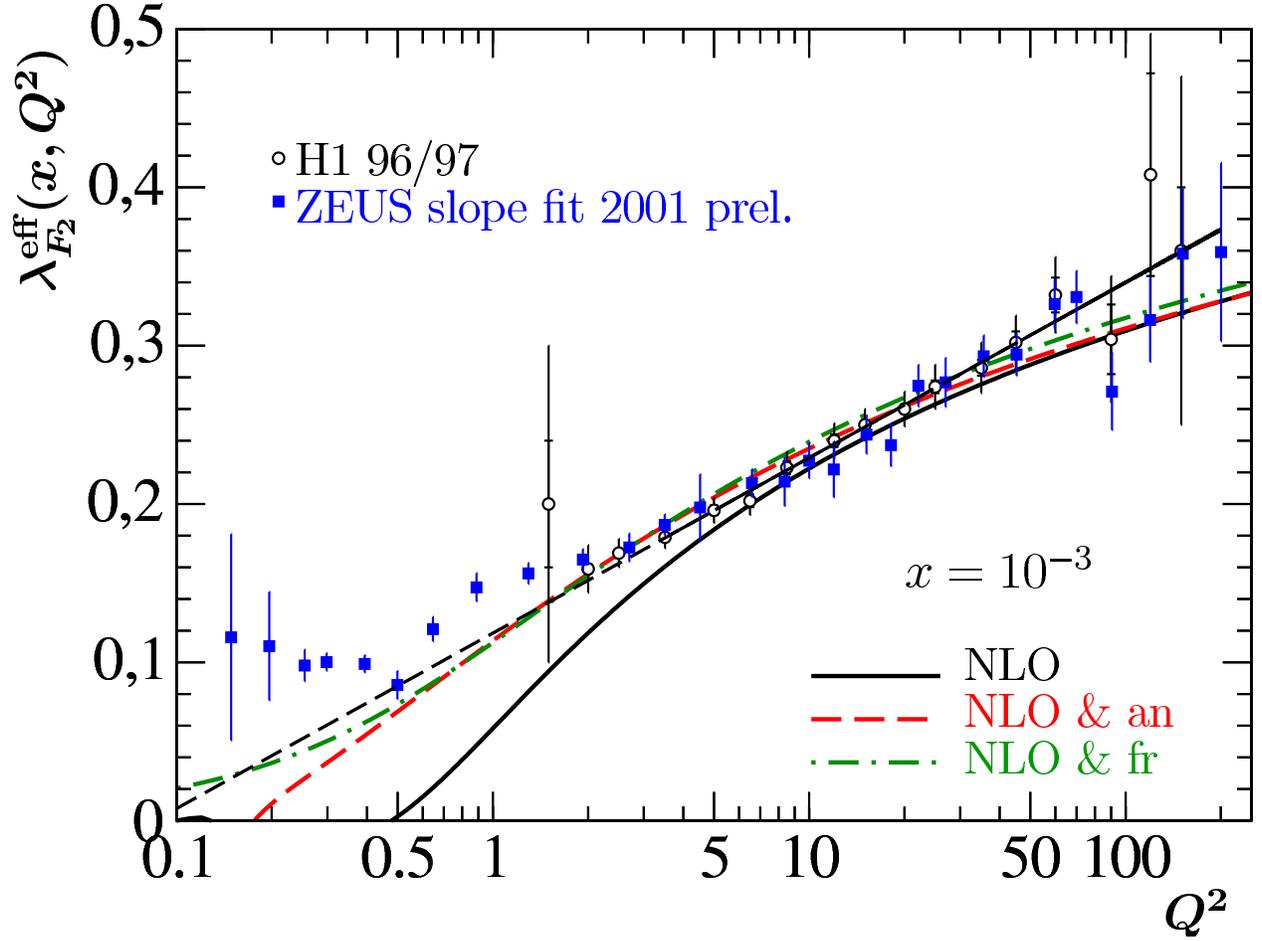}
\end{center}
\vskip 1cm
\caption{The values of effective slope
$\lambda^{\rm eff}_{\rm F_2}$ 
as a function of $Q^2$ for $x=10^{-3}$.  
The experimental points are from H1 \cite{H1slo,DIS02} (open points) and ZEUS 
\cite{Surrow} (solid points). The solid curve represents the NLO fit.
The dash-dotted  and lower dashed curves represent the NLO fits with 
``frozen'' and analytic coupling constants, respectively. The top dashed line
 represents the fit from \cite{H1slo}.}
\label{fig4}
\end{figure}

\begin{figure}[htb]
\begin{center}
\psfig{figure= 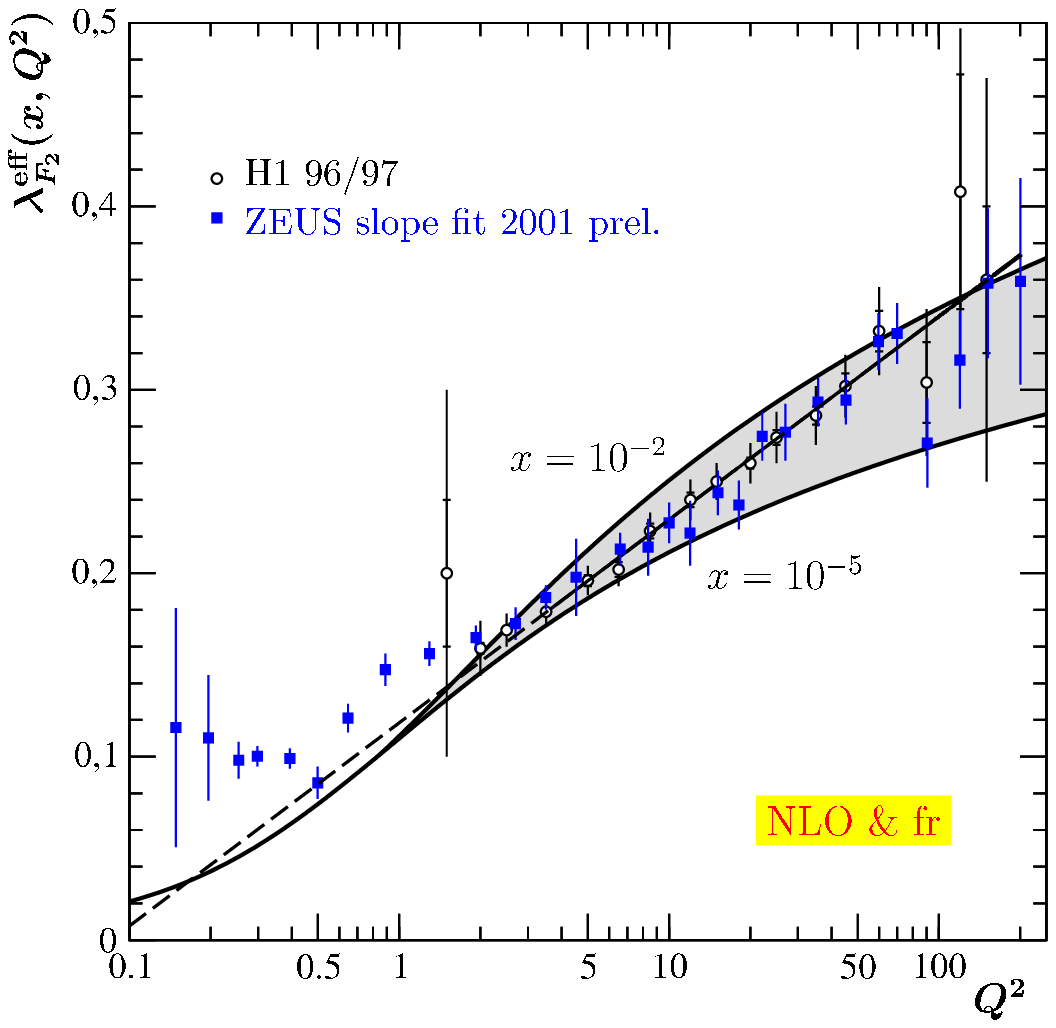,width=17cm,height=13cm}
\end{center}
\vskip 1cm
\caption{The values of effective slope
$\lambda^{\rm eff}_{\rm F_2}$  
as a function of $Q^2$.
The experimental points are same as on Fig. 4.
The dashed line
 represents the fit from \cite{H1slo}.
The solid curves represent the NLO fits with 
``frozen'' coupling constant
at $x=10^{-2}$ and  $x=10^{-5}$.}
\label{fig5}
\end{figure}


\begin{thebibliography}{0}


%
\bibitem{H197} H1 Collab. (C. Adloff {\it et al.}),
Nucl. Phys. B
%{\bf B407} (1993) 515, {\bf B470} (1996) 3, 
{\bf 497}, 3 (1997);
%, Phys. Lett. B {\bf 393}, 453 (1997),
%
%\bibitem{H101} H1 Collab.: C. Adloff {\it et al.},
Eur. Phys. J. C {\bf 21}, 33 (2001).
%%
\bibitem{ZEUS01} ZEUS Collab. (S. Chekanov
%J. Breitweg
%M. Derrick 
{\it et al.}), 
%Phys. Lett. {\bf B316} (1993) 412, 
%Z. Phys. {\bf C72} (1996) 399, {\bf C69} (1996) 607, 
Eur. Phys. J. C
%{\bf 7}, 609 (1999), 
{\bf 21}, 443 (2001).
%,Phys. Lett. {\bf B487} (2000) 53.
%
\bibitem{Surrow} ZEUS Collab. 
(B. Surrow), hep-ph/0201025.
%
\bibitem{H1slo} H1 Collab. (C. Adloff {\it et al.}),
 Phys. Lett. B {\bf 520}, 183 (2001).

\bibitem{DIS02}  H1 Collab. (T. Lastovicka), Acta  Phys. Polon. B {\bf 33}, 
2835 (2002);
 H1 Collab. (J. Gayler), Acta  Phys. Polon. B {\bf 33}, 2841 (2002).
%
\bibitem{CoDeRo} A. M. Cooper-Sarkar, R. C. E. Devenish, and A. De Roeck, 
Int. J. Mod. Phys. A {\bf 13}, 3385 (1998); 
A.~V.~Kotikov,
  %``Deep inelastic scattering: Q**2 dependence of structure functions,''
  Phys.\ Part.\ Nucl.\  {\bf 38}, 1, 828 (Erratum) (2007).
%  [Erratum-ibid.\  {\bf 38}, 828 (2007)].
  %%CITATION = FECAA,38,1;%%
%
\bibitem{DGLAP} V. N. Gribov and L. N. Lipatov, Sov. J. Nucl. Phys. 
{\bf 15}, 438, 675  (1972);
%
%\bibitem{DGLAP}  
L. N. Lipatov, Sov. J. Nucl. Phys. {\bf 20}, 94 (1975);  
G. Altarelli and G. Parisi, Nucl. Phys. B {\bf 126}, 298 (1977);  
Yu. L. Dokshitzer, Sov. Phys. JETP {\bf 46}, 641 (1977).
%
\bibitem{BFKL} L. N. Lipatov, Sov. J. Nucl. Phys.
 {\bf 23}, 338 (1976);
E. A. Kuraev, L. N. Lipatov, and V. S. Fadin, 
 Phys. Lett. B {\bf 60}, 50 (1975); Sov. Phys. JETP {\bf 44}, 443 (1976);
{\bf 45}, 199 (1977); Ya. Ya. Balitzki and  L. N. Lipatov, 
Sov. J. Nucl. Phys. {\bf 28}, 822 (1978); 
L. N. Lipatov, Sov. Phys. JETP {\bf 63}, 904 (1986).
%
\bibitem{fits}  
A. D. Martin and W. S. Stirling, R.G. Roberts,  R.S. Thorne,
Eur. Phys. J. C {\bf 23}, 73 (2002); 
 J. Pumplin {\it et al.} (CTEQ Collab.),  JHEP {\bf 0207}, 012 (2002); 
JHEP {\bf 0602}, 032 (2006).
% Preprint MSU-HEP-011101 (hep-ph/0201195). 
%
\bibitem{GRV} M. Gluck, E. Reya, and  A. Vogt, 
Eur. Phys. J. C {\bf 5}, 461 (1998);
M. Gluck, C. Pisano, and  E. Reya,
Eur. Phys. J. C {\bf 40}, 515 (2005).
\bibitem{Ourfits}
A. V. Kotikov, G. Parente, and J. Sanchez Guillen, 
Z. Phys. C {\bf 58}, 465 (1993);
%%CITATION = ZEPYA,C58,465;%%
G. Parente, A. V. Kotikov, and  V. G. Krivokhizhin,
Phys. Lett. B {\bf 333}, 190 (1994);
%%CITATION = HEP-PH 9405290;%%
A. L. Kataev, A. V. Kotikov, G. Parente, and  A. V. Sidorov,
Phys. Lett. B {\bf 388}, 179 (1996); 
%%CITATION = HEP-PH 9605367;%%
Phys. Lett. B {\bf 417}, 374 (1998);
%%CITATION = HEP-PH 9706534;%%
A. L. Kataev, G. Parente, and A. V. Sidorov,
Nucl. Phys. B {\bf 573}, 405 (2000);
%%CITATION = HEP-PH 9905310;%%
A. V. Kotikov and V. G. Krivokhijine,
Phys. At. Nucl.  {\bf 68}, 1873 (2005)
%%CITATION = PANUE,68,1873;%%
(hep-ph/0108224).
%%CITATION = HEP-PH 0108224;%%

%
\bibitem{BF1}  R. D. Ball and S. Forte,
Phys. Lett. B {\bf 336}, 77 (1994).
%
\bibitem{Munich}  
L. Mankiewicz, A. Saalfeld, and T. Weigl,
Phys. Lett. B {\bf 393}, 175 (1997).
%
\bibitem{Q2evo} A. V. Kotikov and  G. Parente,
Nucl. Phys. B {\bf 549}, 242 (1999); 
%%CITATION = HEP-PH 9807249;%%
Nucl. Phys. (Proc. Suppl.) A {\bf 99}, 196 (2001).
%(hep-ph/0010352).
%%CITATION = HEP-PH 0010352;%%
%{\it in} Proc. of the Int. Conference PQFT98 (1998), Dubna
%(hep-ph/9810223);
%%%CITATION = HEP-PH 9810223;%%
%{\it in} Proc. of the 8th Int. Workshop on Deep Inelastic
%Scattering, DIS 2000 (2000), Liverpool, p. 198 
%(hep-ph/0006197).
%%%CITATION = HEP-PH 0006197;%%
%
%
\bibitem{HT} A. Yu. Illarionov, A. V. Kotikov, and  G. Parente,
%hep-ph/0402173; 
Phys. Part. Nucl. {\bf 39}, 307 (2008);
%%CITATION = HEP-PH 0402173;%%
Nucl. Phys. (Proc. Suppl.) {\bf 146}, 234 (2005).
%%CITATION = NUPHZ,146,234;%%
%
\bibitem{Rujula} A. De R\'ujula, S. L. Glashow, H. D. Politzer {\it et. al.},
%S.B. Treiman, F. Wilczek, A. Zee,  
% et al.,
Phys. Rev. D {\bf 10}, 1649 (1974).
%
%
\bibitem{NMC} NM Collab. (M. Arneodo {\it et al.}),
Phys. Lett. B {\bf 364}, 107  (1995); Nucl. Phys. B {\bf 483}, 3 (1997);
%
E665 Collab. (M. R. Adams {\it et al.}),
Phys. Rev. D {\bf 54}, 3006 (1996);
%
A. Donnachie and P. V. Landshoff,
Nucl. Phys. B {\bf 244}, 322 (1984); {\bf 267}, 690 (1986);
Z. Phys. C {\bf 61}, 139 (1994).
%
\bibitem{lowQ2} ZEUS Collab. (J. Breitweg {\it et al.}),
Phys. Lett. B {\bf 407}, 432 (1997).
%H1 Collab.: C. Adloff et al.,
%{\em Nucl.Phys.} {\bf B497} (1997) 3.
%
\bibitem{lowQ2N} ZEUS Collab. (J. Breitweg {\it et al.}),
Phys. Lett. B {\bf 487}, 53 (2000);
Eur. Phys. J. C {\bf 21}, 443 (2001). 
%
\bibitem{Schrempp} F. Schrempp, hep-ph/0507160.

\bibitem{KoPa02} A. V. Kotikov and G. Parente,
J. Exp. Theor. Phys. {\bf 97}, 859 (2003). 
%%CITATION = HEP-PH 0207276;%%

\bibitem{Greco}
G. Curci, M. Greco, and Y. Srivastava, Phys. Rev. Lett. {\bf 43}, 834 (1979);
                                    Nucl. Phys. B {\bf 159}, 451 (1979);
M. Greco, G. Penso, and Y. Srivastava, Phys. Rev. D {\bf 21}, 2520 (1980);
%M. Greco and the 
PLUTO Collab.(C. Berger {\it et al.}), Phys. Lett. B {\bf 100}, 351 (1981);
%
%\bibitem{NikoZa}
N. N. Nikolaev and B. M. Zakharov,  Z. Phys. C {\bf 49}, 607 (1991);
{\bf 53}, 331 (1992);
%\bibitem{BaKwSt}
B. Badelek, J. Kwiecinski, and  A. Stasto,  Z. Phys. C {\bf 74}, 297 (1997).

%
\bibitem{ShiSo}
D. V. Shirkov and I. L. Solovtsov,  Phys. Rev. Lett {\bf 79}, 1209 (1997);
Theor. Math. Phys. {\bf 120}, 1220 (1999).

\bibitem{LoYn} C. Lopez and F. J. Yndur\'ain, 
 Nucl. Phys. B {\bf 171}, 231 (1980); 
 Nucl. Phys. B {\bf 183}, 157 (1981);
%\bibitem{Yndu} 
C. Lopez, F. Barreiro, and F. J. Yndur\'ain, 
Z. Phys. C {\bf 72}, 561 (1996); K. Adel, F. Barreiro, and F. J. Yndur\'ain,
 Nucl. Phys. B {\bf 495}, 221 (1997).

\bibitem{DoLa} A. Donnachie and P. V. Landshoff,
Phys. Lett. B {\bf 296}, 227 (1992); Phys. Lett. B {\bf 437}, 408 (1998).
%
\bibitem{Abramo}
H. Abramowitz, E. M. Levin, A. Levy, and U. Maor,  
Phys. Lett. B {\bf 269}, 465 (1991);
A. V. Kotikov, Mod. Phys. Lett. A {\bf 11}, 103 (1996); 
%%CITATION = HEP-PH 9504357;%%
Phys. At. Nucl. {\bf 59}, 2137 (1996).
%%CITATION = PANUE,59,2137;%%
%[{\em Yad. Fiz.} {\bf 59} (1996) 2219].



\bibitem{YF93}
A. V.~Kotikov, Phys. At. Nucl. {\bf 56}, 1276 (1993);
%%CITATION = PANUE,56,1276;%%
%[Yad. Fiz. {\bf 56} (1993) N9, 217].
%
%\bibitem{method}A.V. Kotikov, 
%Phys. Atom. Nucl. 
{\bf 57}, 133 (1994);
%%CITATION = PANUE,57,133;%% 
%[Yad. Fiz. {\bf 57} (1994) 142];
Phys. Rev. D {\bf 49}, 5746 (1994).
%%CITATION = PHRVA,D49,5746;%%
%

%
\bibitem{FKR} G. M. Frichter, D. W. McKay, and J. P. Ralston,
Phys. Rev. Lett. {\bf 74}, 1508 (1995).
%
\bibitem{CaKaMeTTV}
A. Capella, A. B. Kaidalov, C. Merino, and J. Tran Thanh Van,  
Phys. Lett. B {\bf 337}, 358 (1994);
A. B. Kaidalov, C. Merino, and D. Pertermann,
Eur. Phys. J. C {\bf 20}, 301 (2001); 
%\bibitem{DeJePa}
P. Desgrolard, L. L. Jenkovszky, and F. Paccanoni,  
Eur. Phys. J. C {\bf 7}, 655 (1999);
%
%\bibitem{VoKoMa} 
V. I. Vovk, A. V. Kotikov, and S. I. Maximov,
Theor. Math. Phys. {\bf 84}, 744 (1990);
%%CITATION = TMPHA,84,744;%%
L.~L.~Jenkovszky, A.~V.~Kotikov, and F.~Paccanoni,
Sov.\ J.\ Nucl.\ Phys.\  {\bf 55}, 1224 (1992);
%%CITATION = SJNCA,55,1224;%%
JETP Lett.\  {\bf 58}, 163 (1993);
%%CITATION = JTPLA,58,163;%%
Phys.\ Lett.\ B {\bf 314}, 421 (1993);
%%CITATION = PHLTA,B314,421;%%.
A. V. Kotikov, S. I. Maximov, and I. S. Parobij,
Theor. Math. Phys. {\bf 111}, 442 (1997).
%%CITATION = TMPHA,111,442;%%
%

\bibitem{MRS}  A. D. Martin, W. S. Stirling, and R. G. Roberts,
Phys. Lett. B {\bf B387}, 419 (1996).

%
\bibitem{FaLi} V. S. Fadin and L. N. Lipatov, 
%hep-ph/9802290.
Phys. Lett. B {\bf 429}, 127(1998);
G. Camici and M. Ciafaloni, Phys. Lett. {\bf B430}, 349 (1998);
%
%\bibitem{KoLi}  
A. V. Kotikov and L. N. Lipatov, Nucl. Phys. B {\bf 582}, 19 (2000);
%%CITATION = HEP-PH 0004008;%%
%Nucl.\ Phys.\ B 
{\bf 661}, 19 (2003).
%%CITATION = HEP-PH 0208220;%%
%
\bibitem{Grunberg}
G. Grunberg, Phys. Rev. D {\bf 29}, 2315 (1984); 
Phys. Lett. B {\bf 95}, 70 (1980).


\bibitem{DoShi}
Yu. L. Dokshitzer and D. V. Shirkov, 
Z. Phys. C {\bf 67}, 449 (1995);
%
%\bibitem{Rsmallx} 
A. V. Kotikov, JETP Lett. {\bf 59}, 1 (1994);
Phys. Lett. B {\bf 338}, 349 (1994);
%%CITATION = PHLTA,B338,349;%%
%
%
%\bibitem{Wong}  
W. K.~Wong, 
Phys.~Rev.~D {\bf 54}, 1094 (1996).
%
\bibitem{bfklp}  S. J. Brodsky, V. S. Fadin, V. T. Kim {\it et al.},
%, L.N. Lipatov, G.B. Pivovarov, 
JETP. Lett. {\bf 70}, 155 (1999);
%\bibitem{Salam} 
M. Ciafaloni, D. Colferai, and G. P. Salam,
Phys. Rev. D {\bf 60}, 114036 (1999); JHEP {\bf 07}, 054 (2000);
R. S. Thorne, Phys. Lett. B {\bf 474}, 372 (2000);
Phys. Rev. D {\bf 60}, 054031 (1999); {\bf 64}, 074005 (2001); 
G. Altarelli, R. D. Ball, and S. Forte,
Nucl. Phys. B {\bf 621}, 359 (2002).
%
\bibitem{Andersson}  Bo Andersson {\it et al.}, 
Eur.\ Phys.\ J.\ C {\bf 25}, 77 (2002).
%(hep-ph/0204115).
%%CITATION = HEP-PH 0204115;%%

%
\bibitem{Nesterenko}
A. V. Nesterenko, Phys. Rev. D {\bf 64}, 116009  (2001);
Int. J. Mod. Phys. {\bf A18}, 5475  (2003);
A.~V.~Nesterenko  and J.~Papavassiliou,
 %``The massive analytic invariant charge in QCD,''
 Phys.\ Rev.\  D {\bf 71}, 016009  (2005);
J.\ Phys.\ G {\bf 32}, 1025 (2006);
G. Cvetic, C. Valenzuela, and I. Schmidt, 
%hep-ph/0508101;
Nucl. Phys. Proc. Suppl. {\bf 164}, 308 (2007);
G. Cvetic and C. Valenzuela, J. Phys. G {\bf 32}, L27 (2006);
%hep-ph/0601050;
 Phys.\ Rev.\  D {\bf 74}, 114030  (2006);
Phys.\ Rev.\  D {\bf 77}, 074021  (2008);
 A. P. Bakulev, S. V. Mikhailov, and N. G. Stefanis,
Phys. Rev. D {\bf 72}, 074014 (2005);
Phys.\ Rev.\  D {\bf 75}, 056005  (2007);
 R.~S.~Pasechnik, D.~V.~Shirkov, and O.~V.~Teryaev,
  %``Bjorken Sum Rule and pQCD frontier on the move,''
  Phys.\ Rev.\  D {\bf 78}, 071902  (2008);
 R.~S.~Pasechnik, D.~V.~Shirkov, O.~V.~Teryaev, O. P. Solovtsova, and
V. L. Khandramai,  arXiv:0911.3297 [hep-ph].

%
\bibitem{Cvetic}
G. Cvetic and C. Valenzuela,
Braz.\ J.\ Phys.\  {\bf 38}, 371 (2008);
 A. P. Bakulev, S. V. Mikhailov, arXiv:0803.3013 [hep-ph];
N. G. Stefanis, arXiv:0902.4805 [hep-ph].

\bibitem{Cvetic1}
G. Cvetic, 
%{\it et al.},
A.Yu. Illarionov, B.A. Kniehl, and A.V. Kotikov,
 Phys. Lett. {\bf B679}, 350 (2009).
%%CITATION = PHLTA,B679,350;%%

%
\bibitem{KoLiZo}  A. V. Kotikov, A. V. Lipatov, and N. P. Zotov, 
 J.\ Exp.\ Theor.\ Phys.\  {\bf 101}, 811 (2005).
%  [Zh.\ Eksp.\ Teor.\ Fiz.\  {\bf 128}, 938 (2005)]
%  [arXiv:hep-ph/0403135].
  %%CITATION = HEP-PH 0403135;%%



\end{thebibliography}
\end{document}